\documentclass{article}   
\usepackage[authordate]{biblatex-chicago}
\usepackage{physics,amsmath,amssymb,mathrsfs}
\usepackage{enumitem}
\usepackage{hyperref}

\title{Spacetime emergence: an (in)effective story}
\author{Mike D. Schneider}
\date{[Accepted at \emph{Philosophy of Physics}]}

\begin{document}
\maketitle

\begin{abstract}
Physicists and philosophers are increasingly prone to regarding our current physical theories as providing `effective descriptions' of real-world systems. In the context of quantum gravity research, this fuels a common view that the classical spacetime theory of general relativity provides effective descriptions where it is successfully applied. That common view of general relativity, in turn, encourages an `effective' understanding of spacetime emergence. But descriptions of spacetime in general relativity irreducibly include global physical content, which is not effective. Recognizing this fact reigns in the interpretive scope of the common view of general relativity and specifically undermines our thinking about spacetime emergence effectively.

\end{abstract}

\section{Introduction}

Some theoretical physicists are interested in the relationship between descriptions of dynamical systems according to our current best theories and fundamental, non-spatiotemporal descriptions unknown. So are many philosophers, under the banner of `spacetime emergence'.

Spacetime emergence is particularly intriguing in light of our modeling physical systems that possibly differ from one another in their global spacetime structure, according to our current best theories. Examples include singularities in Big Bang cosmology and in analytic studies of black holes \parencite{earman1995bangs}; evaporating black holes in semi-classical gravity \parencite{manchak2018information,lesourd2018causal}; possibly spatially closed models of the evolution of large-scale structure \parencite{coles1997universe}; and even, arguably, (non-vanishing) cosmological constants in the asymptotic regimes of isolated systems \parencite{Belotgoodtimesroll,schneider2022empty}, as well as inertial versus accelerating thermometers confined to the `Rindler wedge' in the context of standard particle theory \parencite{earman2011unruh}.\footnote{There are also more `exotic' applications of our current physics where global spacetime structure matters. These range in plausibility from, e.g., the old mainstream proposal that topological defects might serve as seeds of structure formation \parencite{smeenk2018inflation} to the science fiction hypothesis that time machines may be possible \parencite{earman2009laws}.} 

The relevance of these diverse examples, taken altogether, to ongoing fundamental physics research, and particularly to quantum gravity, spoils any means of thinking about the global features of spacetime as something externally fixed, e.g. by brute empirical fact or as a Humean `Best Systems' choice.\footnote{I raise a caveat to this claim at the end of the Conclusion section.} Rather, it is plausible that we may (sometimes) be interested in thinking about underlying \emph{physical explanations} of the `global physical content' that features alongside the local in a given spacetime model. One may formulate this interest as a live research question: wherever spacetime is thought to emerge in an application of some underlying theory to some fitting worldly target, what in the underlying theoretical description is meant to especially account for the emergence of the spacetime's global physical content?\footnote{An illustration, which begins with an amusing aside. Freeman \citeauthor{dyson1979time} (\citeyear{dyson1979time}) once pondered the ambiguity involved in proclaiming our large-scale universe to be spatially open rather than closed. Admitting that he found the closed possibility claustrophobic, he then proposed, tongue in cheek, the ultimate project in climate engineering or existential risk management: \begin{quotation}Supposing that we discover the universe to be naturally closed and doomed to collapse, is it conceivable that by intelligent intervention, converting matter into radiation and causing energy to flow purposefully on a cosmic scale, we could break open a closed universe and change the topology of space-time so that only a part of it would collapse and another part of it would expand forever? I do not know the answer to this question. If it turns out that the universe is closed, we shall still have about $10^{10}$ years to explore the possibility of a technological fix that would burst it open. (p. 448)  \end{quotation} Set aside the interesting conundrum \citeauthor{dyson1979time} raises about how to get emergent topology \emph{change} out of an underlying quantum theory that is unitary. Presently, empirical observations via standard model cosmology favor a spatially open model (I imagine \citeauthor{dyson1979time} relieved, before his recent passing). But we have no grasp as to why. That is, forty-five years since \citeauthor{dyson1979time}'s remarks, we still do not know what could underwrite the basic fact that: in an assumed globally hyperbolic application of the classical theory of general relativity to a suitably uniform quantum cosmos (so: excluding topology change by fiat), the relevant spacetime one ought to consider when setting up the standard model is one whose Cauchy surfaces are open, not closed.} 

In this article, I argue that this research question is overlooked on an `effective' understanding of spacetime emergence that is encouraged by one common view of general relativity (GR), a classical spacetime theory of gravity. The primary purpose of the article is therefore to spotlight something that is otherwise at risk of being neglected in philosophers' excitement to lean into the common view. Indeed, I provide one recent example of this neglect in the context of foundations of string theory in \S\ref{secStrings}. 

According to the common view I have in mind, by associating gravitational states with spacetimes, GR singles out spacetime geometry as providing an `effective description' of the dynamics of a gravitating system. Here, `effective description' is a term of art and endearment, which is inspired by (i.e. moves beyond) the formalism of effective field theory (EFT). Once that term of art is clarified, the basic observation of the article is that assumptions about global physical content are necessary to vindicate the common view. So wherever the common view is applicable, this is because we feel entitled to certain modeling resources within the application at hand, which are not themselves explained as part of an effective description of gravity. But where we are entitled to taking those resources for granted, we are not in the position to pursue satisfying answers to the research question posed above about whence the global physical content of spacetime within that application.

Explicating the term of art `effective description' is the focus of \S\ref{secEffective}. But as I intend it, the following is true. Theoretical physicists, as well as increasingly many philosophers, are generally accustomed to regarding our current best theories as providing `effective descriptions', whether or not the formalism of EFT strictly applies. The common view of GR is therefore an instance of a wider trend. Still, specific to GR, there are two (partially coincident) arguments to support the common view. Both arguments are developed in \S\ref{sectionCommonView}. 

This article is not an objection to the common view of GR. It is an objection to taking an understanding of spacetime emergence motivated by that view as the end-all, concerning spacetime emergence in applications of quantum gravity. The crux of the objection is this. I will argue in \S\ref{secGlobalSpacetimeStructure} that details about global spacetime structure relevant in describing a gravitating system in GR act like \emph{superselection sectors} for its possible classical gravitational field states.\footnote{Might this be more than analogy? Perhaps, but proceed with caution. The application of formal arguments available in quantization theory about pre-quantum structures in the classical theory that correspond with superselection sectors in the quantum theory may not obviously extend to the case of scalar field theories with infinite degrees of freedom (let alone to tensor field theories or metric theories like GR). I thank Ben Feintzeig for some correspondence on this point.} This observation curtails the interpretive scope of the common view of GR, in such a way as to undermine thinking `effectively' about spacetime emergence. Namely, thinking of GR as providing effective descriptions of gravitational states for applicable target systems in quantum gravity is a \emph{post-superselection interpretation} of the relevant spacetime geometry (on either of the two arguments supporting the common view). The interpretation is only justified, provided that `superselection rules' have been independently supplied (i.e. with respect to whatever is the given application at hand, given specifics of that application). But recall the research question about what in an underlying theory could account for emergent spacetime's global physical content. If we simply assume the superselection sector in the classical theory is given (say, in the setup of the problem that motivates application of the classical theory to the quantum gravity system at hand), we have lost the ability to furnish interesting answers to that question. In other words, explanations of the superselection rules that are needed to justify, in any particular application of quantum gravity, the suitability of the common view of GR must also figure centrally in our telling of the emergence of the relevant spacetime. 

\section{\label{secEffective}Effective Descriptions}

What precisely physicists and philosophers mean by `effective' is rarely clear. But a nice entry point is provided by \citeauthor{huggett1995renormalisation} (\citeyear{huggett1995renormalisation}):\begin{quotation} [...] imagine that you have a [quantum field theory] with lots of different particles with widely varying masses. In situations where there isn't enough energy to create a certain particle, parts of the theory which make reference to that particle can be ignored. Thus at such low energies, physics is described by an [EFT] which `effectively' captures everything relevant. (p. 171-172)\end{quotation} `Effective' has something to do with descriptions provided by EFTs, as are relevant to describing quantum fields specifically within suitably low-energy regimes. (In fact, EFTs have also found uses in other modeling contexts, including hydrodynamics and in the classical astrophysics of compact binaries. But enthusiasm for EFTs comes from their uses in condensed matter and particle physics, both of which concern quantum physics.) Meanwhile, it is common to treat our current best physical theories \emph{as if they were EFTs}: ``[EFTs] are almost everywhere, and everything is considered to be [one]'' \parencite[p. 267]{hartmann2001effective}.

But not everything in our current best physics is given as a quantum field theory (QFT). So, not all physical systems are (currently) best considered as quantum fields. (And nor are those other systems obviously better handled by reformulation into EFTs about some other objects besides quantum fields.) How then are we to understand the sentiment expressed by \citeauthor{hartmann2001effective}?

Plausibly, the sentiment being expressed is that we commonly treat our current best physical descriptions of real-world systems as \emph{effective descriptions}, whether or not the surrounding theory is an EFT. This amounts to coining the italicized phrase as a new term of art. ``Art'' because 1) the underlying idea must be imprecise, so that treating everything as an effective description falls short of an outrageously ambitious technical conjecture (namely: that all relevant physical theories can and will, at the end of the day, be adequately reformulated as EFTs); and 2) few who think about everything in physics as an effective description seem committed to seriously pursuing the status of such a conjecture, even in restricted domains. So the language invoked must be more like metaphor.\footnote{What is more: the phrase is not meant to entail any commitment to a logical or linguistic theory semantics (despite what a use of the word ``description'' might hint at). I thank a reviewer for recommending I explicitly disavow the reading otherwise.}

Here is the gist. Empirical evidence for the wide, successful application of a dynamical theory in fundamental physics delimits a `familiar' descriptive regime, in the sense of warranting a certain kind of proposition: that the application of the theory `effectively' captures everything relevant about the system being so modeled, (just) so far as concerns all of its \emph{known} dynamical behavior --- all of which is documented as falling within the familiar regime. Thus, one concludes that the theory provides effective descriptions of all relevant physics: when assessed only within that familiar regime, the system behaves precisely as we know it to behave. Its dynamics are precisely as the theory dictates. 

The emphasis here is on `regime' talk, in the absence of underlying physical theory suitable to license that talk. This is in contrast to the situation within the quoted passage from \citeauthor{huggett1995renormalisation}, where the familiar regime is a restriction of a given QFT to just one of its ``parts'' --- namely, a low energy part. More exactly, the quotation from \citeauthor{huggett1995renormalisation} concerns `top-down' EFT: wherever a QFT is taken to aptly model a real-world system as a UV complete (i.e. renormalized) quantum field, a top-down EFT may be constructed that effectively captures everything relevant about that quantum field below some specified energy scale. In essence, an explicit statement of the EFT encodes all influence of the high energy (`UV') physics of the real-world system on the low-energy dynamics, \emph{given that the system is considered as a quantum field within the originating QFT, ranging over all scales}. The EFT is thus a way of physically modeling a (low-energy) part of the quantum field within that QFT as autonomous or quasi-decoupled from all other parts \parencite{Bain2013} --- essentially, as a quantum field with dynamics all its own. This autonomy is especially pronounced on a `full-group' interpretation of the renormalization group flow that relates the QFT to the EFT describing the quantum field's low energy part.\footnote{For discussion of the full-group interpretation, see \parencite{KOBERINSKI202314}, who argue for its importance in high enery particle physics contra a `semi-group' interpretation more reasonably deployed in condensed matter physics.} On a full-group interpretation, the EFT enjoys equal footing with the QFT, rather than being cast as its subsidiary. 

In the case of `bottom-up' EFT, things are a bit different. Instead of deriving an autonomous expression of a low-energy part of a UV complete quantum field, one rather seeks an autonomous expression of the physics that \emph{would be relevant} within applications of \emph{some or other} QFT, should one only wish to restrict attention to a part of the UV complete quantum field (referenced within that some or other QFT), which happens to fall below a specified energy scale. That is, bottom-up EFT makes reference to a low-energy regime within an unspecified QFT --- an instance of regime talk without an explicit underlying theory suitable to license it. 

As stated, bottom-up EFT seems idiosyncratic, perhaps unmotivated. But if one is working with QFTs, particularly those that are non-renormalizable, the upshot of bottom-up EFT is a toolkit for interpreting quantum fields already in use: they provide descriptions that effectively capture everything relevant within low-energy regimes. Such is an appealing interpretation, in that it allows us to avoid committing ourselves to too much about what the same systems are like within as-yet unfamiliar high-energy regimes --- or indeed, to the question ``regimes of what?''.\footnote{\label{fnStandardEFT} There are two qualifications to be made here. First, \citeauthor{koberinskiSmeenkRethinkingCCP} (\citeyear{koberinskiSmeenkRethinkingCCP}) argue that the unrestricted use of bottom-up EFT as a toolkit for interpretation leads to absurdity in certain modeling contexts within our current theoretical physics, e.g. the (semiclassical) cosmological sector. That there may be such theoretical limitations to our use of this toolkit can support speculations about how to study novel physics beyond the scope of EFTs \parencite{koberinski2024breakdown}. Second, the `swampland' program considers limitations with the unrestricted use of this toolkit even for interpreting quantum field theories thought to represent a primary use case, in light of recent developments in quantum gravity research \parencite{palti2019swampland}. Here, the idea is that perhaps not every quantum field theory can be considered a theory of the low-energy, effective part of a system ultimately well described by a UV complete quantum theory that includes gravity. So, in present context, the swampland program pursues qualifications to the claim that the bottom-up EFT toolkit evades the question ``regimes of what?'' in worlds anything like our own.} That is, given a QFT that is (known to be) empirically supported below some energy scale in application to some or other system, the descriptions of that system provided by the QFT can be regarded as capturing everything relevant about it \emph{because} we are confining our attention exclusively to matters of fact about the quantum field within an accessible, familiar regime. 

It is this bottom line of the bottom-up EFT as an interpretive toolkit for empirically supported QFTs, which serves as inspiration for the new term of art `effective description'. Empirically supported applications of a physical theory yield effective descriptions of the relevant real-world systems in that they capture everything relevant within a familiar regime. We can safely regard a part of the system's full dynamics --- whatever those dynamics ultimately are --- as of no further importance, given our restricted focus. Where successful then, what has happened in the production of an effective description is that one imagines a bulk system of unknown dynamical composition to have been \emph{partitioned}. (Note: the partitions are drawn within a theoretically-motivated abstract space, not spacetime.) So there are parts to the system's dynamics, which we may proceed to associate with bulk subsystems that are event-wise coincident in spacetime, yet where at least one of whose dynamical evolution is autonomous. That one subsystem is exactly as described by the theory providing the effective description, with any relevant constants fixed as fit parameters by empirical means in the usual course of theory construction. On this view, for instance, the reduction of the full dynamics of GR to the Friedmann equations in application to the classical study of the `zero-order' universe provides an effective description of the general relativistic cosmos as uniformly expanding space --- consistent with common talk on the subject, where a perturbation theory is then called upon to model the growth of large-scale structure amidst that expansion \parencite{dodelson2003modern}.\footnote{Some researchers have attempted to build an explicit EFT of large-scale structure (e.g. \parencite{carrasco2012effective}). This is a different matter from thinking about standard model cosmology as providing an effective description. The virtue of attempting to build an EFT of large-scale structure is to subsequently assess it as a competitor to the approach taken in the standard model. That is, in the terms used presently, it allows one to formulate a comparative question: which approach --- standard model cosmology or cosmology as EFT hydrodynamics --- provides a more/better effective description of the evolution of large-scale structure, given our empirical record?}

Hence: effective descriptions capture everything relevant (i.e. `are descriptions') of a system's bulk dynamical evolution confined to within a familiar regime. And statements that delimit those confines --- cutoff scales, in a more general sense than the cutoff in an EFT --- are statements about an autonomous subsystem's \emph{boundary}. Some precedence for this usage of the technical term `boundary' may be found in \parencite{wilson2017physics,bursten2021function}. Both of these authors provide extended discussions meant to generalize the notion of boundaries for some bulk material (e.g. the concrete walls of one versus another physical box, which each hold taut a string that reverberates in the interior according to a shared, i.e. `universal', simple harmonic law) to cover a whole motley of expository details necessary to successfully model physical systems as autonomous, in terms of applications of dynamics understood to otherwise govern universally over systems that share a description in the bulk.

\section{A Common View of GR}\label{sectionCommonView}

We were just led to a term of art, `effective description', in an effort to make sense of a sentiment: that we (or, many of us) think of everything in our current best physics as \emph{basically like} an EFT, even non- quantum field theories like standard model cosmology. In the context of quantum gravity research, this general inclination has resulted in a view that GR, a classical spacetime theory of gravity, ultimately provides effective descriptions of \emph{gravitating systems}: fundamental physical systems for which the strength of the gravitational interaction cannot be safely ignored, e.g. those total systems --- universes --- that are of interest within the study of quantum cosmology. So although we anticipate that a future theory of quantum gravity will force radical revisions to the descriptions we take as fundamental of those systems, including our universe as a whole, it is nonetheless common to think that the dynamics governing spacetime geometry in the context of GR live on, in that they provide an effective description. There is some way of partitioning the system, where the familiar part is general relativistic spacetime as we are already accustomed to describing, and meanwhile any and all departures from what we are accustomed get safely sequestered in a different part --- as-yet unknown dynamics sequestered within an as-yet unfamiliar regime. 

This is as far as it seems to me one can get, from an embrace of the term of art `effective description' alone. But in order for it to be useful \emph{that} we can think of GR as providing an effective description --- that there is an autonomous subsystem, which is adequately described by the tools of GR --- one needs to specify something of the boundary of the regime regarded as familiar, or the beginning of the unfamiliar. 

One natural thought coming from EFTs is to look for a high-energy cutoff scale. Another natural thought coming from the breakdown of the Friedmann equations in expanding universe cosmology is to look for a microscopic cutoff scale. Fortunately, these two thoughts pair well: given a choice of frame on the spacetime, which sets frequency (hence, energy) scales for assessing the system, it is sensible to track regions about an event that are suitably small that quantum corrections to spacetime geometry would be on or above order unity. The largest of those regions yields the `Planck regime' about that event --- a microscopic, utterly unfamiliar domain in which only quantum gravity presides \parencite{callender2001physics}. The construction to have in mind here is that of solid ball centered on the event, with radius given by a fiducial cell within which quantum corrections to the metric at that event, which are due to high-energy physics, cannot be assumed to be suppressed.\footnote{As stated, this construction also applies to spacetime singularity resolution, in the sense of regions containing blow-ups of curvature invariants. In these cases, regardless of the high-energy physics of matter fields in the vicinity, one expects high-energy contributions in the gravity sector to contribute quantum corrections, under the reasoning that new theory will circumvent all such blow-ups (cf. \parencite{crowther2022four}). Note, also, that spacetimes in general lack suitable asymptotics to define a frame at all; this discussion implicitly assumes some or other quasi-local notion of energy.} Within the solid ball, the theoretical terms from GR --- those describing spacetime geometry --- are not apt; without the solid ball, all is as familiar: GR. 

Solid balls drawn about events are not the only place to look for the beginning of the unfamiliar. One expects the breakdown of GR as well, where fluctuations in the matter fields contained within an arbitrarily large spacetime region are allowed to grow large. But working with that part of the boundary of GR's effective dominion requires a dynamical theory of gravity coupled to quantum matter, which relates to GR in a controlled manner. (We will see one such theory, shortly.) Stochastic gravity is a general phenomenological framework for studying beyond the breakdown of GR in this latter sense, by means of the `Einstein-Langevin equation' \parencite{hu2008stochastic}. In this setting, two-point correlations in the (semiclassical) stress-energy tensor operator contribute a noise term that back-reacts on spacetime geometry. (In principle, higher order correlations ought also to contribute.) Unlike the cutoff relevant to demarcating the Planck regime \emph{about an event}, in this case, one imagines specifying a noise cutoff, equivalent to demanding that two-point correlations be sufficiently suppressed \emph{globally in spacetime}. Only beyond that cutoff are quantum fluctuations in the matter fields considered descriptively relevant. Below, all is as familiar: GR.

To my knowledge, stochastic gravity has not received serious philosophical study. Its significance in present context is merely to underscore how weak is the term of art `effective description', deployed in an interpretation of GR within its empirically supported applications. It is possible that greater attention to the noise cutoff in that formalism would lead to better understanding the thermodynamic or hydrodynamic aspects of spacetime emergence. As it stands, I gather that a more ordinary assumption is that the common view of GR includes exclusive focus on the exit from the Planck regime, defined event-wise in spacetime (and given a frame on the spacetime), as the premier aspect of the boundary of effective description. The explanation for this assumption seems to be a social-historical one that has to do with the rise of the research sub-field of high-energy particle physics and its emphasis on scattering events, rather than anything substantive: the exit from the Planck regime is, essentially, a high-energy cutoff imposed on the classical gravitational theory, beyond which it does not make sense to model particle scattering.

I said in the Introduction that there are two arguments to support the common view of GR. I now qualify that remark: there are two arguments to support the common view of GR, \emph{supplemented by our focusing narrowly on there being a high-energy cutoff to the effective description, specified event-wise in spacetime}. Both arguments proceed from general reasoning about gravity, as represented in GR. The first proceeds by interpreting GR in terms of a classical relativistic field theory of gravity; the second engages in similar reasoning, but instead makes use of a stipulated relationship between classical gravity within GR and a quantum gravitational field within a low-energy EFT that \citeauthor{burgess2004quantum} and \citeauthor{WallaceLowEnergyQGDraft} have each discussed under the name `low-energy quantum gravity'. (Low-energy quantum gravity includes multiple regimes that are unfamiliar from the vantage point of GR, including regimes characterized by large fluctuations in the stress-energy, in which some of the formalism of stochastic gravity can be recovered.) Low-energy quantum gravity is widely regarded as both empirically supported and physically significant in the context of ongoing quantum gravity research, so stipulating a relationship between it and GR is plausible. On the other hand, to the extent that one might wish to remain neutral on the nature of the coupling of gravity to quantum matter in the absence of crucial experiments,\footnote{\label{fnBet}Jonathan Oppenheim  has entered into a public bet with Carlo Rovelli and Geoff Penington at 5000/1 odds that such crucial experiments (e.g. of gravitationally mediated entanglement predicted by the quantum theory, critically discussed in \parencite{huggett2023quantum}) will fail to confirm that spacetime is described by a quantum theory. Indeed, depending on the experimental protocol and background theory, the terms of the bet are also met if certain experiments carried out succeed in confirming that spacetime is classical --- cf. \parencite{oppenheim2023gravitationally}). For a signed copy of the terms of the bet, see \url{https://www.ucl.ac.uk/oppenheim/pub/quantum_vs_classical_bet.pdf} (last accessed 30 September, 2024).} the first argument stands where the second has little bite.

\subsection{Two arguments in support of the common view}\label{subsecArguments}

Above, GR was referred to as a classical spacetime theory of gravity. In greater detail: GR associates with gravitational interaction events a certain classical relativistic dynamics of the spacetime metric, sometimes regarded in terms of a local physical law throughout space and time (understood as the total arena over which a field theory of matter is defined). The theory admits of description in terms of a Cauchy problem, whether in the presence or absence of matter. It also admits of description in terms of a local action principle, given a pre-geometric, smooth spacetime manifold and additional boundary conditions. We thus say that locally, GR is a classical theory governing the spacetime metric, qua \emph{relativistic gravitational field} interacting with classical (relativistic) matter.\footnote{ \parencite{wald1984general} provides textbook presentations of both of these descriptions of the theory (see appendix E in the case of the latter).}

Local event-wise pictures of gravity in GR, like the two just summarized, are quite fruitful. Consider the opening remarks in the widely influential treatise by \citeauthor{hawking1973large} (\citeyear{hawking1973large}), which describes the scope of their mission: 
\begin{quotation}[...] we shall take the local physical laws that have been experimentally determined, and shall see what these laws imply about the large scale structure of the universe.

There is of course a large extrapolation in the assumption that the physical laws one determines in the laboratory should apply at other points of space-time where conditions may be very different [...] In fact most of our results will be independent of the detailed nature of the physical laws, but will merely involve certain general properties such as the description of space-time by a pseudo-Riemannian geometry and the positive definiteness of energy density. (p. 1)
\end{quotation}
In other words: there is a whole descriptive science concerning space and time as are relevant to a gravitating universe, which is entailed by treatments of GR that associate spacetime geometry with the realization, everywhere and when, of some local physical law having to do with gravity. 

\subsubsection*{GR in terms of a spin-2 theory of gravity}

One may push this point further, which brings us into the first of the promised arguments. In GR, given any two states of the local gravitational field that agree on the underlying manifold structure $M$ (i.e. understood as two metric tensor fields defined on $M$), one can always find a third state, also understood as defined on $M$, such that the first is the (tensorial) sum of the second and third. So, for any state $g$ and preferred `background' state $g_0$, one can linearly decompose $g$ into a `background' $g_0$ and `foreground' $(g-g_0)$.\footnote{This feature of the theory immediately follows from the fact that the metric is represented as a tensor and tensors are multilinear. Also, note that the dynamics of GR, given by Einstein's equation, are non-linear, and so it is generally not the case that if $g_0$ and $(g-g_0)$ are solutions to Einstein's equation (for some matter fields, or else in vacuum), so is $g$.} It is this feature of the state space for the gravitational field on a fixed smooth manifold, per GR, which paves the way for the construction of a field theory of the classical `graviton': an expression of the gravitational field on $M$, defined against a choice of absolute spacetime background diffeomorphic to $M$. In the notation above, the field theory of the classical graviton would proceed by treating $(g-g_0)$ to some perturbative order on an absolute spacetime background $(M,g_0)$. If $(M,g_0)$ is chosen so as to be isotropic, one may even rigorously define a massless spin-2 graviton `particle' that is fit for this classical field theory \parencite[ch. 9]{huggett2021out}.\footnote{For a textbook treatment of the graviton concept, see \parencite[ch. 2.1]{kieferquantumgravity2ndedition}. Also, it may be that $M$ admits no metrics that are isotropic. I set this issue aside.} Fixing such a background (perhaps as a matter of convention), one is tempted to reinterpret GR as a theory that ascribes to a gravitating system, event-wise on a manifold, a local interacting classical physics of the relativistic gravitational field. This is to entertain understanding GR, qua spacetime theory, through its reduction to a `spin-2 theory' of gravity as a dynamical, universally coupled relativistic field on a spacetime background \parencite{petrov2020field,salimkhani2020dynamical,Salimkhani2024}.

Such a `spin-2 interpretation' of gravity in GR has recently been criticized by \citeauthor{linnemann2023gr} (\citeyear{linnemann2023gr}). The authors object to the view that a spin-2 theory of gravity reduces GR, as the standardly claimed derivation of GR from basic commitments about a massless spin-2 graviton is anything but. Their focus assumes $g_0$ is flat, above and beyond isotropic; but the criticism goes through more generally. Indeed, that the spin-2 interpretation is instantiated in multiple different choices of non-isometric (in fact, non-diffeomorphic), isotropic conventions for $g_0$ figures in passing as part of their argument, which is structured in terms of two horns: ``either the view is physically incomplete in so far as it requires recourse to GR after all, \emph{or it leads to an absurd multiplication of alternative viewpoints on GR}'' (from the abstract, emphasis added).\footnote{\citeauthor{Salimkhani2024} (\citeyear[p. 132]{Salimkhani2024}) has since criticized the criticism, basically arguing that the two horns are each blunt. There can be a reduction (and an argument showing it), even if the reasons one would ever employ to take the steps involved in that reduction have to do with making recourse to GR (first horn), and meanwhile metatheoretical considerations (e.g. having to do with achieving good explanations, commitments to systematicity across fundamental physics, and so on) can get involved to pare down the ``absurd multiplication of alternative viewpoints'' (second horn). I am somewhat sympathetic to the latter point, and I see myself in the following paragraph as embracing what is ultimately a similar argument, in defense of thinking about GR as a theory that uses spacetime geometry to describe a fundamental gravitational field. On the other hand, the main purpose of this article is to articulate dissatisfaction with taking such a view of GR as the final word, essentially because those same metatheoretical considerations, when they get applied case by case to vindicate this view of modeling by means of GR, obscure lingering questions about whence the global physical content of the relevant spacetime model.}

Nonetheless, as \citeauthor{linnemann2023gr} recognize, this view of GR has been very influential in the particle physics community's thinking about the fundamental physics of gravity, despite its shortcomings. So even granting that reduction is too strong of an intertheoretic relationship between the spin-2 theory and GR, the former theory may still be helpful in that it provides a (partial) interpretation of the latter. At minimum, it plausibly motivates thinking about GR as a theory that uses spacetime geometry in certain contexts to provide, event-wise, descriptions of some \emph{fundamental gravitational field}. Those descriptions cannot be exact of that fundamental field (for all the remarks made about the Planck regime above), so they must be effective descriptions of the fundamental field considered outside of the Planck regime. Perhaps a theory of that fundamental field would also implicate the otherwise fixed spacetime background as an effective description. 

It is noteworthy that this view does seem to permit a scenario whereby gravity is ultimately proven classical and stochastic (rather than quantum) within the Planck regime, in which case GR could be understood to provide an effective description of a classical, fundamentally stochastic field (cf. the ``postquantum classical theory of gravity'' proposed in \parencite{oppenheim2023postquantum}). This is arguably a virtue of the present defense of the common view, especially compared to the other defense coming next. Still, as the possibility that gravity is fundamentally classical and stochastic is generally considered a fringe possibility only not outright precluded by tabletop experiment (see footnote \ref{fnBet}), I largely set it aside for the remainder of my argument and treat the two defenses with indifference.

\subsubsection*{GR in terms of low-energy quantum gravity}

As in the first argument, in defense of the common view of GR, we may begin from general reflection on the status of the gravitational field within the theory. Accordingly, spacetime geometry captures, everywhere and when there is such a field, local gravitational physics. In the first argument, GR was explicitly recast as a theory that provides effective descriptions of a fundamental gravitational field without the Planck regime, by means of that spacetime geometry. A connection was made between the interpretive toolkit used in that recasting and certain attitudes in the particle physics community. 
 
Instead of interpreting GR in terms of another classical theory on the way to quantum gravity, one might, however, be inclined to skip the middle-man (thereby excluding the basic possibility --- noted at the end of the first argument --- that gravity ultimately proves to be fundamentally classical and stochastic). Why? Consider that, although it amounted to a crucial development in the dynamics of theory development in early twentieth-century physics, through a contemporary lens, GR could appear a non-starter. As we have come to understand the quantum nature of matter, we have equally come to rely on theoretical innovations beyond GR in explanations of the stability of various exotic systems of interest in geo- and astrophysics, many of which are now squarely within empirical reach. Those innovations would seem chiefly to involve treating the gravitational field as a quantum field locally interacting, event-wise, with all of the interacting quantum fields implicated in standard model particle physics (though there are, perhaps, workarounds that could save the relevant explanations case by case, should gravity prove not quantum --- ergo, the bet referenced in footnote \ref{fnBet}). Hence, as argued by \citeauthor{WallaceLowEnergyQGDraft} (\citeyear{WallaceLowEnergyQGDraft}), the collection of innovations --- gathered up under the umbrella of low-energy quantum gravity --- forms a part of our current best physics, in that its regime of applicability is familiar and available for empirical study, with ample confirmation bestowed. Moreover, it subsumes the equivalently familiar domain of empirically-supported GR (setting aside, for the sake of exposition, issues raised by black holes). And as a QFT (albeit non-renormalizable), we can in turn interpret low-energy quantum gravity as a low level in an attractive `tower of EFTs' picture of particle physics: successive EFTs adequately screening off higher-energy physics all the way up.

So, perhaps best is to interpret GR as \emph{replaced by} low-energy quantum gravity (though, see cautionary remarks in footnote \ref{fnStandardEFT}). If so, where GR is empirically adequate, that adequacy is explained by its providing effective descriptions of the local physics of low-energy quantum gravity in the right restricted settings: where low-energy quantum gravity is itself taken as descriptively apt about an interaction event (so, not in the Planck regime about that event) and, moreover, where fluctuations of all kinds are sufficiently suppressed.\footnote{For more details on low-energy quantum gravity, see \parencite{burgess2004quantum}.} 

On this view, GR provides effective descriptions insofar as it effectively captures everything relevant to the mean field physics of the quantum gravitational field, referenced in applications of low-energy quantum gravity to interaction events where the influence of gravity contributes significantly. 

\subsection{An effective understanding of spacetime emergence}\label{subsecEventWiseEmergence}

The common view that GR provides effective descriptions of a gravitational field, given in the context of either of the two arguments just rehearsed, fosters an effective understanding of spacetime emergence. According to the effective understanding, spacetime is what one gets as one zooms out from a gravitational interaction event (and provided one can ignore effects of matter being quantum). Recall the solid ball construction about an event represented spatiotemporally in GR: within the solid ball, we do not take spatiotemporal language as descriptively apt. But beyond the surface of that ball, GR may be understood to provide accurate descriptions about the same event. The picture is that of a `spacetime foam', introduced originally by \citeauthor{wheeler1955geons} and later championed by Hawking: where standard (smooth, etc.) spacetime modeling techniques, e.g. identifications/constructions of worldlines and clocks in large-scale cosmology, black hole physics, and even terrestrial particle physics, are taken as apt whenever the ratio of the Planck scale to the curvature scale of the spacetime about a point (and also the ratio of the Planck scale to the scale of the relevant excitations in all of the interacting quantum fields at that point) nearly vanishes. Otherwise, and in particular within the Planck regime where that condition is thought to fail, those standard techniques are to be discarded for something radically new.

Thus, as one moves in a sequence from arbitrarily high-energy scales to low-energy scales about a fixed event, there is a transition whereafter effective descriptions of the underlying gravitational dynamics approximately correspond to a spacetime description provided by GR. Roughly, this is the scale over which one fails to resolve the exit from the Planck regime about an observer as distinct from the event in which the observer passes. We hereby arrive at an event-wise picture of the emergence of a classical spacetime description of a gravitating system, e.g. our universe as a whole: classical spacetime emerges as one shifts down to units that fail to resolve the Planck regime about an event (relative to an observer) as distinct from the event itself.

The catch in all of this, as I will now explore, is that this view of spacetime emergence requires that a local field theory of spacetime geometry (qua gravitational field) exhausts the physical content of GR. Yet, except when specific details of a given modeling situation constrain us in the boundary, this misses out on the \emph{global} physical content implied by the use of the classical theory.

\section{Global Spacetime Structure}\label{secGlobalSpacetimeStructure}

We just saw two arguments for the common view of GR, and the event-wise, effective understanding of spacetime emergence encouraged by it. Crucially, the common view focuses attention on the `local' content of GR, or a view of GR as a local theory providing effective descriptions of some gravitational field (whether a more fundamental gravitational field that could be classical or just a stipulated quantum gravitational field).  

But as stressed in the Introduction, global spacetime structure matters in our modeling a variety of real-world systems, according to our current best physics. Indeed, the quoted passage at the beginning of section \S\ref{subsecArguments}, which is taken from the beginning of \parencite{hawking1973large}, is a setup for a whole treatise exploring what, arguably, \emph{cannot be concluded} about the large scale structure of spacetime just from thinking about GR in terms of a descriptive science of the gravitational field. This is a theme picked up at length by \citeauthor{earman1995bangs} (\citeyear{earman1995bangs}), and more recently by \citeauthor{manchak2020global} (\citeyear{manchak2020global}).

In this section, I argue that this well-known and basic fact about GR --- that it is not merely a local classical field theory of some relativistic gravitational field, or else a stand in for low-energy quantum gravity --- significantly curtails the view that GR provides effective descriptions, and so also undermines our thinking effectively about spacetime emergence. Namely, in GR, states of gravity are global spacetime geometries that we might associate with a gravitating system. It is only having first fixed the diffeomorphism class of the spacetime that we are then able to think about gravitational field configurations. And even then, we must be cautious. The causal structure of singular spacetimes can differ wildly from the causal structure of non-singular spacetimes in the same diffeomorphism class. Meanwhile, de Sitter spacetime is diffeomorphic to the Einstein static universe, but only the former has cosmological horizons (and neither is singular). And as demonstrated by  G{\"o}del spacetime, it need not be that, by continuous local adjustments consistent with the local dynamics of GR, one may move from a gravitational field configuration without time travel on a manifold to one with time travel \parencite{stein1970paradoxical}.\footnote{\label{fnStein} I take the general moral here to be that global spacetime structure is (necessarily) more than just topology or underlying manifold structure/diffeomorphism class. What more is, however, an open question. From the above considerations (and perhaps motivated by thinking about Penrose diagrams), one might propose that global spacetime structure equates conformal structure. But this won't do: a number of conformal properties are not stable, and in the context of quantum gravity it is reasonable to assume that quantum corrections can cause lightcones to `wiggle'.} In a nutshell, spacetimes have ``global'' spacetime structure, or details that are necessary in order to use GR to study gravitating systems in terms of the dynamics of an associated gravitational field. And meanwhile, we lack clarity regarding what to cleave off of a spacetime as that ``global'' part. 

Or, to put the point differently: the field observables in GR are only defined after \emph{at least} (but probably not only!) fixing the global manifold structure relevant in a given application. This should put us in the mindset of thinking about global spacetime structure as akin to a superselection rule in the classical theory, applied to various systems. Hence, the theoretical structure of GR includes some global physical content (nebulously defined), which might itself be understood as fixing something of the boundary, in suitable applications of the bulk dynamics of the theory to particular gravitating systems. Here, I am again using the term ``boundary'' in the sense clarified at the end of section \ref{secEffective}: global physical content provides descriptive ingredients relevant for dynamically modeling gravity in the bulk by means of GR.\footnote{See \parencite{rovelli1991observable} for a classic foundational discussion about observables in GR, in relation to a fixed diffeomorphism class. Putting his terminology in conversation with mine to be introduced presently, superselection rules would be provided in a given application by appeal to some external object relevant there. Denying such objects generally suggests that we are not, in general, able to regard gravitational states as (diffeomorphism invariant) gravitational field states, as the common view of GR would have it. For more on the issues with observables in quantum gravity versus quantum observables given the common view, see \parencite{giddings2006observables}.} 

How does this standard observation about the theoretical structure of GR impact the two arguments for the common view of GR? Or, to put the point in the opposite manner: where go the superselection rules in our setting up a relationship between GR and either a classical spin-2 theory or low-energy quantum gravity? Clearly they are already assumed: in the first case, it is the global structure of the background $(M,g_0)$, assumed to be isotropic (if not, moreover, flat); in the second case, it is the global structure taken to be relevant in the mean-field description of the quantum gravitational field. Both arguments interpret GR through one or another post-superselection framework for thinking about the problem of quantum gravity: the superselection rules are absorbed into the description of the absolute spacetime background on which a fundamental, interacting gravitational field lives, or else in how one specifies the global physical content involved in stating the mean-field approximation of the quantum gravitational field.

In other words, I do not dispute the relevance of thinking about gravity in terms of a relativistic field dynamics applied event-wise in the study of gravitational interactions, and nor do I dispute low-energy quantum gravity as providing viable descriptions of (by theoretical supposition) real-world, gravitating quantum systems. What I would like to emphasize, however, is that the descriptions of the relevant gravitating system provided in each of these cases concern (possibly quantized) gravitational \emph{field} states, not (possibly quantized) spacetime geometries. I say that I am emphasizing this point; to be sure, it is not a new point in the context of quantum cosmology. Here, the upshot is that spin-2 gravity and low-energy quantum gravity are theoretical frameworks for thinking effectively about GR, after the application of superselection rules that fix, at least, the diffeomorphism class of the spacetime, necessary to render gravitational states in GR as gravitational field states. Fine. But absent independent support for some given superselection rules in any particular case of a gravitating system, this is not a wholesale view of the success of GR, qua spacetime theory, in application to that system. The outstanding question for research is, rather: what might that articulated support look like, within the underlying gravitational theory? 

But this question is easily obscured if we think of spacetime emergence effectively. This is because the effective understanding of spacetime emergence proceeds from an assumption that details of the application at hand, which concern the description of the real-world gravitating system to be supplied by the future theory of quantum gravity, already include suitable superselection rules relevant in the familiar descriptive regime without the Planck regime. Given that assumption, one may indeed consider spacetime geometry in GR as providing an effective description of what \emph{remains} of the gravitational dynamics (i.e. having interpreted the superselection rule as a feature of the boundary for modeling an autonomous subystem as a classical gravitational field). But in general, absent such independent support for superselection rules that would recover the requisite field theoretic thinking about gravitational states, there is little justification for thinking that classical spacetime is an effective description of the underlying dynamics of the real-world gravitating system. In other words, it is generally inappropriate to regard the classical description as an instance of effective emergence of spacetime upon an exit from the Planck regime.

\section{A Stringy Perspective}\label{secStrings}

\citeauthor{huggett2021out} (\citeyear{huggett2021out}) provide a detailed, foundational account of the effective emergence of spacetime within the context of a string theory approach to quantum gravity research. (The presentation also doubles as a convenient foothold into the general subject, with further references found within.) Specifically, they show how well-controlled, stringy physics in a perturbative scattering regime can give rise to ``the observable open set structure of spacetime'' (p. 12) alongside its metric structure about a scattering event, and in doing both, they claim to unpack a fairly standard conclusion in string theory that ``... GR is an effective, phenomenal theory describing the collective dynamics of strings (in certain quantum states) in target space'' (p. 8). It will be fruitful here to reflect briefly on their account, as at a glance it is easy to miss the question of global physical content of GR within their conclusion. So, one may get the impression that the details of string theory vindicate thinking about spacetime emergence effectively --- at least, as long as a string theory approach to quantum gravity research remains viable. 

Though prima facie reasonable, I argue that this impression is unfounded. Just as in the case of the argument using low-energy quantum gravity, in order to think about spacetime emergence effectively in the context of string theory, one requires something akin to a superselection rule, which settles possible differences in global spacetime structure as a matter of boundary. The enriched physical story provided in the string theory approach just further specifies this point: one requires constraints on the viable (global) spacetime geometry of the fundamental `target space' in which the strings (perturbatively) live, from which one computes an effective `phenomenal' spacetime description by means of a sum over string scattering diagrams, or paths through that target space.\footnote{For analysis of `target' versus `phenomenal' space, see \parencite{huggett2017target}).} The oversight by \citeauthor{huggett2021out} stems from their article's (reasonable) focus on establishing the claim that the definite spatial localization of the general relativistic metric in phenomenal spacetime --- that is, its \emph{local} open set structure about a scattering event --- is genuinely emergent: some bit of physical description newly relevant to the description of the event as occurs in phenomenal spacetime, i.e. not already something evident of the stringy description of the same event, which would (all the same) seem to include something like spatial localization of the event in target space. Their ultimate argument to this effect is to draw on the existence of dualities to underscore that the topology of target space is ``in some sense mere formal representation, and the open set structure of spacetime must arise from some other string theoretic structure that is invariant under dualities (p. 15).'' 

This argument proceeds by demonstration using the case of T-duality.\footnote{This is a strong choice for demonstration; as they note, other dualities in fact ``do more violence'' (fn. 10 in the chapter) to the topological structure in one model, mapped over to its dual.} In particular, in the context of T-duality, they conclude that the open set structure of spacetime about a scattering event in phenomenal spacetime emerges from some common core of T-dual string scattering models, which either capture the relevant physics as localized stringy interactions occurring in an open region of target space (in one theory) or as non- spatially localized stringy interactions occurring in a closed, cylindrical region in an associated `winding space' (in the dual theory).

Granting, for the sake of argument, their ancillary views about what amounts to physical content in T-dual models,\footnote{Dualities within and without string theory have been studied widely by philosophers of physics (e.g. those articles gathered in the special issue ``Dualities in Physics'' (\citeyear{DualitiesinPhysicsSHPS}) within \emph{Studies in History and Philosophy of Modern Physics}). A ``common core'' view of the physical content of theories related by duality, endorsed by \citeauthor{huggett2021out}, is just one of many, and it is not without drawbacks (e.g. \parencite{grimmer2024duality}).} I have no qualms with their claims on this point: agreed, because of T-dualities, it is inappropriate to regard the spatial localization of scattering events in phenomenal spacetime as inherited from a definite physical fact about localization of the underlying event in target space. Intuitively: the dual description in winding space must accommodate all of the same phenomenal physics, and clearly in that description the stringy physics is not localized in an open region of target space. So there remains some further fact that bears responsibility for localization in the phenomenal spacetime, which only in one dual can possibly be stated as a fact about localization in target space. Hence, the open set structure emerges from something in the common core, which only looks like localization of stringy scattering on one non-unique choice of description. Meanwhile, dynamics well-described by string scattering diagrams in target space neatly give rise to the general relativistic metric about the scattering event in phenomenal spacetime (and an equivalent dynamical account goes through in winding space in the T-dual model), so, the general relativistic metric about the localized event emerges thereupon from certain (or other, dual) kinds of string excitations. Hence, we find the emergence of relativistic spacetime geometry out of perturbative string theory. 

But this does not equate the emergence of relativistic spacetime, in the sense of global manifold structure with metric familiar in GR. There are two points to be made here. First, \citeauthor{huggett2021out}'s argument is silent as to the relationship between phenomenal spacetime and, fixing some or other choice of dual description of the underlying stringy physics, possibly non-trivial global structure in target space (or in whatever other space, the topology of which is meant to gives rise, in that choice of dual, to the open set structure of phenomenal spacetime). The basic point is that we are working in perturbative string theory, and while one hopes that the ``stupendous'' cut-and-paste constructions \parencite[p. 59]{butterfield2014under} familiar from the foundations of classical GR may be ignored as possible `background fields' structuring target space --- that a ``strings all the way down'' picture (i.e. M-theory) will force us into more comfortable waters --- there is no guarantee. Again, none of this conflicts with \citeauthor{huggett2021out}'s argument: the suggestion is not to deny the emergence of sufficient topological structure to ``localize'' the scattering in phenomenal spacetime, but to deny that this story suffices, in the context of quantum cosmology, to account for the emergence of general relativistic spacetime, with all of its attendant global physical content.

This last comment leads to the second point. The methods of perturbative string theory evidently involve restricting to scattering events that may, in phenomenal spacetime, be considered to occur somewhere an S-matrix can be defined: typically, regions that we might imagine properly embedding in spacetimes with asymptotically flat spacetime structure, and for which it is reasonable to ignore the possibility that the regions include points coincident with a globally defined horizon.\footnote{For the sake of exposition, I am setting aside hopes in string theory motivated by a conjectured AdS/CFT duality that horizons in a bulk setting can be dealt with holographically, with reference to a boundary S-matrix determined by the CFT.} In other words, the application of this particular spacetime emergence story in the context of quantum cosmology is limited: supposing that the emergence arguments provided by \citeauthor{huggett2021out} quantify over every member of some given, unstructured collection of phenomenal events, to ensure that the suite of methods called upon is apt, the emergent open set structure that results is only the (global) topology of the phenomenal spacetime if certain facts about global structure are already specified. Without that additional specification, the relativistic spacetime geometry recovered is merely the structure that is preserved among the members of some collection of spacetimes that are locally isometric in the sense of \citeauthor{manchak2020global} (\citeyear{manchak2020global}) --- enough to recover local properties of spacetime (including a metric, everywhere!), using \citeauthor{manchak2020global}'s definitions, but not global properties.

In all particular applications, one might just as well assume the necessary added details are provided by modeling context. That is, in addition to the global structure of the small compactified spatial dimensions in string theory, which are what stand to give rise to familiar particle physics in the phenomenomal spacetime (if only we could find the correct structure!), so too may we assume that the global structure of the large dimensions of the phenomenal spacetime are picked out by further contextual facts, spelled out in the underlying stringy models. But the ultimate explanation for the global structure of the large phenomenal spacetime dimensions --- why the context of application provides the relevant constraints on the global spacetime geometry of the large dimensions --- awaits us as a story in M-theory. The success prospects of an effective understanding of spacetime emergence therefore rest, in a stringy approach, on an additional explanation of the emergence of spacetime's global physical content in M-theory. With sufficient care, then, the specifics of formalism encouraged on a string theory approach to quantum gravity do not seem to add much at present, so far as concerns the scope of the argument advanced in this article. 

\section{Conclusion}

I have argued that a common view of GR in quantum gravity research --- namely, as providing effective descriptions of gravitating systems --- is inadequate, barring independently motivated superselection rules that would render an application of GR suitable for description in terms of a gravitational field. This is not an objection to the common view, but it is an objection to a view of spacetime emergence that goes with the common view. We cannot understand spacetime emergence effectively, as spacetime emergence involves emergence of spacetime's global physical content. Yet global physical content is already accounted for, if we assume a superselection sector as given, in keeping with the common view of GR. So effective spacetime emergence is circular: it misses altogether an answer to the question of what accounts for the global physical content of the emergent spacetime.

This condemnation may be too harshly put. Recently, \citeauthor{jaksland2023many} (2023) have argued in the philosophy of quantum gravity for increased specificity as to whence the emergence of a variety of separable ``aspects'' of spacetime, under different proposals in quantum gravity research. From the perspective advanced in their article, it is easy to summarize the argument that has been made here in a more conciliatory tone. We need (also) attend to the emergence of global physical content relevant in classical gravitational descriptions of a given dynamical target. Recovering, in a suitable approximation scheme, even all of the local physical content of our familiar models of general relativity as an effective description of underlying non-spatiotemporal physics is not the whole story. So, for instance, in the foundations of string theory discussion of \S\ref{secStrings}, we can recognize --- per \parencite{huggett2021out} --- that perturbative string theory gives sufficient resources to generally establish, in a string theory approach to quantum gravity, the emergence of some physics consistent with our modeling certain gravitating systems by means of the dynamics ascribed to spacetime within GR. But the same resources fall short of establishing the emergence of the global physical content of the spacetime, which inevitably appears in any use of GR to describe that physics. That latter aspect of the emergence of spacetime in a string theory approach, overlooked in \parencite{huggett2021out}, remains unexplained.

The claim has been made that this aspect of spacetime emergence --- the question of whence the emergence of global physical content --- is easily neglected in light of the common view of GR. But attention is a precious resource, and the claim has not yet been made that such neglect is to any real detriment. There are two points to be made here, which will together wrap up the discussion. First, I am aware that some readers who have made it this far likely feel underwhelmed by the argument that has occurred. As one reviewer nicely put the point: proponents of an effective view of GR are likely happy to admit that their view minimizes the importance of global features of spacetime. And as far as pills to swallow go, this one may even be medicinal: some proponents of an effective view might be proponents \emph{because} minimizing the importance of global features of spacetime in GR facilitates a hope that the future theory of quantum gravity might just suffice to put an end to all fuss about global spacetime structure raised by those applications of current physics, which motivated us in the Introduction. (Some researchers have expressed this hope to me in private conversation and correspondence.) Such proponents would simply object to the claim that we ought to attend to any aspect of spacetime emergence that has us take seriously global physical content within GR. 

In fact, I am sympathetic to this objection, which is essentially one about how mathematical details of our current physics are to be taken as speculative leads in our pursuit of future physics. As I see it, I have advanced a prima facie reasonable argument that, inasmuch as our going theory of gravity makes use of global physical content (see: those applications in the Introduction), we need to attend to questions in ongoing research about whence that global physical content; and a certain view on the theory makes it easy to neglect such questions. The onus is on those who object to that argument, along the lines just described, to defend that it is unproblematic, in the course of ongoing quantum gravity research, to embrace a view that minimizes the importance of global features of spacetime. One common reaction I have encountered is that global features of spacetime are somehow beyond the empirical reach of the physical theory. So perhaps the defense to be formulated is that non-global features of spacetime all but exhaust the empirical content of the theory. Of course, distinguishing empirical content from theoretical content of a physical theory is famously some trouble. But an accompanying difficulty in the present case would be to formalize a notion of ``global features'' of the theory, to show that we can subsequently dispense with all global features wherever we are only interested in capturing GR's empirical content. The most sophisticated effort I can imagine along these lines would be to borrow \citeauthor{manchak2020global}'s (\citeyear{manchak2020global}) definition of local spacetime properties (mentioned briefly in \S\ref{secStrings}) to formulate a principle: only physically interpret local spacetime properties, not global properties. If one could write down an empirically adequate theory in terms of just local spacetime properties, or at least gesture to that effect, a major step will have occurred in our understanding of global structure in GR, especially as relates to an empirically adequate, future theory of quantum gravity. But hopefully it is clear that this is a substantive project, such that its success cannot be assumed from the start. Equally, one imagines that such a thorough defense of minimizing the importance of global physical content in GR may turn up new resources for the interlocutor to reply, or even new discoveries in the physics that shrink the distance between the contrasted views. 

In the meantime, there is the second point to be made that is specifically in defense of a claim that neglect of global physical content in our thinking about spacetime emergence is detrimental. Namely, because of at least some of those applications of our current physics mentioned in the Introduction, the topic of emergent global physical content is already implicated in two of the leading efforts within contemporary research in quantum gravity phenomenology --- Big Bang singularity resolution and black hole evaporation.\footnote{It is implicated as well, though I am unaware of any discussion explicitly on the point, in phenomenological models of quasi-de Sitter expansion --- interpreted as a mechanism for inflation or dark energy --- as excitations in a quantum gravity theory with (low-energy) anti-de Sitter ground state. But in the context of quantum gravity phenomenology in general, the restriction to theories of quantum gravity formulated with an anti-de Sitter ground state makes this further case parochial.} In the case of Big Bang singularity resolution, one can ask whether or not an adequate resolution to the Big Bang singularity is one that is meant to affix another (prior) spacetime epoch to our presently expanding cosmos \parencite{thebault2023big,schneider2023efforts}. Different answers would seem to disagree in some way about the global physical content of the emergent cosmos (e.g. its asymptotic structure). In the case of black hole evaporation, one paradoxical feature of black hole evaporation concerns the global spacetime properties that seem fitting in our semiclassical models of the process \parencite{manchak2018information}. To the extent that one would like the future quantum gravity theory apt for the Planck regime to resolve all of the paradoxes that seem to crop up in research surrounding semiclassical black hole evaporation \parencite{schneider2022role}, more work is needed on how intertheoretic relations might fix details about which of those global spacetime properties are, ultimately, inappropriate to assume in the semiclassical models. On the quantum gravity end of those intertheory relations would seem to lie explanations within the underlying theory for the emergence of the global physical content of whatever spacetimes ultimately crop up for use at the other end.

A takeaway here is that we would perhaps do well to think of global spacetime geometry as a topic of general import within quantum gravity phenomenology, an area of research focused on model building within various approaches or proposals to develop a theory of quantum gravity. But this takeaway is inherently speculative, and the route for making progress upon taking it is not particularly clear.\footnote{I am cautiously optimistic about drawing morals by analogy from the vast sea of models explored in condensed matter theory, which relate theories across scales always relative to particular applications/target systems. I thank a reviewer for reminding me of this optimism in the context of the present argument.} An alternative speculation is that quantum gravity is exclusively a ``bulk'' theory: essentially incapable of furnishing explanations for global spacetime geometry.\footnote{\label{fncaveat}Here, I am taking for granted that quantum gravity does not have its own global spacetime geometry built in. While it is conceivable that a future theory may have such structure, I think it is safe to say that such a future theory would not count as quantum gravity in the sense ordinarily meant today (insofar as work on spacetime emergence is common across all standard research approaches).} In this scenario, applications of GR do not emerge within applications of quantum gravity; rather, models of quantum gravity provide a fine-grained study of what is dynamically possible within a familiar, macroscopic spatiotemporal world that is otherwise given. 

Is this scenario conceivable? I do not know. But there are hints in the direction. In \parencite[p. 20]{huggett2021out}, we find a functionalism according to which quantum gravity approaches earn their physical content through relations the relevant quantum gravity theories bear to familiar, macroscopic spacetime: ``Rather than following the Canberra plan of vindicating spacetime objects by reduction, in our approach to [quantum gravity] things are reversed: the non-spatiotemporal objects
of [quantum gravity] are vindicated via their identifications with spatiotemporal objects.'' The view assumes an ontology within which it makes sense to talk about spatiotemporal objects concurrently with quantum gravity dynamics, which recalls the possibility dismissed in the Introduction that global features of spacetime might ultimately be regarded as fixed by brute empirical fact in the context of quantum gravity research. That is: contrary to what was supposed in the Introduction, global features of (familiar, macroscopic) spacetime may be taken as fundamental --- hence, they are stipulated --- while nonetheless occupying a different ``level'' than the objects of quantum gravity (which are also considered fundamental).

My sense is that such a view, which limits the scope of quantum gravity to a microscopic bulk description of what is possible about states of affairs of matter confined within a classical and macroscopic spacetime, is radical within the context of quantum gravity research. But it may be viable. And if it is, with quantum gravity dynamics and macroscopic spacetime structure held on par, quantum gravity research starts to look more like a project dedicated to improving multiscale modeling. This has two principal effects. First, it puts phenomenology first: work on developing quantum gravity theory is now intended foremost as a means for locating techniques suited for the instrumental improvement of models of familiar (multiscale) physical targets. Second, it changes how we might think about the work there is to be done within quantum gravity phenomenology. Namely: in multiscale modeling, e.g. in condensed matter physics and materials science, the phenomenology to be explained is inherently multiscale. And one recognizes an ineliminable role played by handshakes between theoretical descriptions that are each suitable for limited descriptions of the same physical system articulated at different scales, each one alone recognized as inadequate to capture the relevant phenomena. 

Here, then, may be a natural place to close: a fork in the road, or two competing speculations in quantum gravity research that follow our attending to what is otherwise an easily neglected aspect of spacetime emergence (given the commonness of the common view of GR).  Either we ought to consider global spacetime geometry a topic of import within quantum gravity phenomenology, where modeling choices in GR are somehow to be teased out of modeling commitments made given the underlying quantum gravity theory and the relationship GR bears to that theory, or else we ought to radically reorient ourselves as to the nature of the work to be done relating developments in quantum gravity theory to those physical systems under study in quantum gravity phenomenology. 

\section*{Acknowledgements and Funding}

Research for this article began while I was a postdoctoral researcher on the Beyond Spacetime project, funded by a `Cosmology Beyond Spacetime' grant (no. 61387) from the John Templeton Foundation. I am grateful for frequent conversations with Nick Huggett at the time, which motivated me to produce the specific argument contained in this article. Adam Koberinski provided a nice sanity check at a much later point, once I had picked up the project again. And I benefited greatly from my anonymous reviewers in the final stages. 

\clearpage
\printbibliography

\end{document}